\documentclass[review]{elsarticle}
\usepackage{amsmath}
\usepackage{supertabular}
\usepackage{graphicx}
\usepackage{epstopdf}
\journal{Journal of \LaTeX\ Templates}

\begin{document}

\begin{frontmatter}

\title{Vaccination dilemma on an evolving social network}
\author{Yuting Wei$^1$}
\author{Yaosen Lin$^2$}
\author{Bin Wu$^1$\corref{mycorrespondingauthor}}
\cortext[mycorrespondingauthor]{Corresponding author (bin.wu@bupt.edu.cn)}
\address{$^1$ School of Sciences, Beijing University of Posts and Telecommunications, 100876, China}
\address{$^2$ School of Information and Communication Engineering, Beijing University of Posts and Telecommunications, 100876, China}

\begin{abstract}
Vaccination is crucial for the control of epidemics.
Yet it is a social dilemma since non-vaccinators can benefit from the herd immunity created by the vaccinators.
Thus the optimum vaccination level is not reached via voluntary vaccination at times.
Intensive studies incorporate social networks to study vaccination behavior,
and it is shown that vaccination can be promoted on some networks.
The underlying network, however, is often assumed to be static,
neglecting the dynamical nature of social networks.
We investigate the vaccination behavior on dynamical social networks using both simulations and mean-field approximations.
We find that
the more robust the vaccinator-infected-non-vaccinator links are
or the more fragile the vaccinator-healthy-non-vaccinator links are,
the higher the final vaccination level is.
This result is true for arbitrary rationality.
Furthermore, we show that, under strong selection,
the vaccination level can be higher than that in the well-mixed population.
In addition, we show that vaccination on evolving social network
is equivalent to the vaccination in well mixed population with a rescaled basic reproductive ratio.
Our results highlight the dynamical nature of social network on the vaccination behavior,
and can be insightful for the epidemic control.
\end{abstract}


\end{frontmatter}


\section{Introduction}
The control of influenza has been important for human beings.
Vaccination has been the principal strategy for the control of infectious diseases since \cite{ref26,ref27} it's invented.
It takes time and money for vaccinators to take vaccination.
Vaccinators are free from disease with a large likelyhood.
For simplicity, we assume that the vaccination is perfect,
i.e., whoever takes the vaccination would be free from disease \cite{ref10,ref11}.
Yet side effects would also occur from time to time including fever.
When the vaccination level is sufficiently high,
the unvaccinated individuals benefit from the herd immunity created by vaccinated individuals.
Consequently the unvaccinated are not likely to get infected.
In other words, herd immunity is similar to the 'public goods' in the tragedy of commons \cite{ref2,ref3,ref4,ref5,ref7,ref8},
where public goods are collected from those who contribute.
It naturally results in a social dilemma between vaccinators and unvaccinators.

Therefore, vaccination is a social dilemma \cite{ref2,ref5,ref11,ref58}.
Vaccinated individuals remain healthy.
For unvaccinated individuals, there are two cases: some of them benefit from herd immunity and remain healthy without paying anything, i.e., successful hitchhikers; the rest get infected and bear a treatment cost.
Therefore, self-interested individuals try to avoid vaccination and benefit from herd immunity.
Such hitchhiking leads to a low vaccination level,
which can be lower than the optimum vaccination level required for the population.
Consequently, it leads to a failure to eradicate the disease.
Evolutionary game theory has been widely adopted in the study of vaccination,
because individuals' welfare in the vaccination is not only up to their decisions to take vaccination but also is determined by others' decisions to take vaccination \cite{ref10,ref11,ref17,ref28,ref29}.

There are two stages in the vaccination.
One is vaccination campaign and the other is epidemic spreading.
Vaccination campaign typically occurs before epidemic outbreak.
At this time, it is typically assume that individuals vaccinate voluntarily.
In other words, individuals adjust their decisions to take vaccination or not.
The unvaccinated individuals are likely to get infected
if they interact most often with other unvaccinated individuals.
Vaccinators are free from disease.
When the epidemic season comes,
disease spreads via contacts.
There are many models of disease transmission,
such as SIS \cite{ref48}, SIR \cite{ref5,ref11,ref41}, SEIR \cite{ref51}, SEIQR \cite{ref52}, SAIS model \cite{ref53} and so on.

When the epidemic season is over,
individuals try to adjust their decisions to vaccinate or not.
Typically imitation rule is adopted: individuals compare their own payoffs with the others to determine whether to vaccinate for the next outbreak of influenza.
Influenza viruses mutate so often that every year the influenza needs an alternative vaccine.
Therefore, we also assume that no one is immune to the mutated virus unless it takes new vaccine.
Previous works study vaccination campaign on a static population structure,
such as scale-free network, random graph network, square lattice and so on \cite{ref5,ref11,ref30,ref43,ref45,ref48,ref49}.
It is found that the hub nodes in static network are likely to take vaccination,
which it is crucial to inhibit the outbreak of epidemic on a degree heterogenous network \cite{ref48}.
This is because the more neighbors there are, the higher the infection possibility   is.
At the same time, hub nodes' vaccination is beneficial to the improvement of the herd immunity for their neighbors.
It highlights the importance of vaccination when heterogenous network structure  is taken into account to model epidemic spreading.
In addition,
it also suggests that it can be effective to  control epidemics via deactivating links in a static network \cite{ref49}.
However, social relationships among individuals varies from time to time.
It is shown that the epidemics could be controlled via social adjustments \cite{ref13}.
This intrinsic nature of social network has been neglected in modelling vaccination, to the best of our knowledge.
On the other hand, interestingly,
dynamical networks have been intensively studied in the field of epidemic spreading and cooperation, respectively.
In epidemic spreading, susceptible individuals are likely to stay away from the infected.
In cooperative social dilemmas, cooperators are prone to connect with other cooperators to get rid of the exploitation by defectors.
Similarly, unvaccinated individuals tend to keep away from those who are unvaccinated,
and try to connect with vaccinators to be protected from the herd immunity.
In other words, herein unvaccinators try to connect more often with the vaccinated.
We explicitly model the dynamical nature of the social network.
And we study the vaccination campaign on a stochastic dynamical network,
where the individuals adjust not only their strategies to vaccinate or not to but also their social relationships \cite{ref23,ref50,ref55,ref57}.

\section{Model}
\begin{figure}[!ht]
	\centering
	\includegraphics[width=1\linewidth]{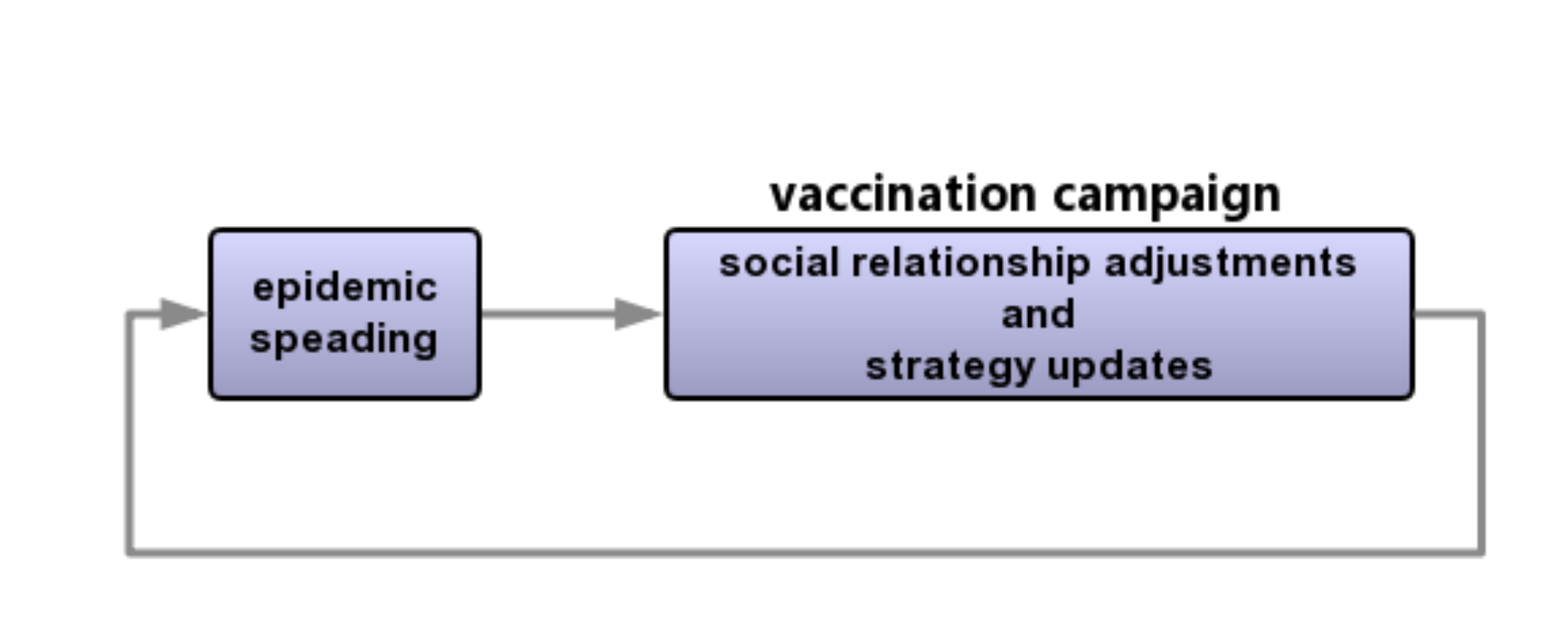}
	\caption{\textbf{Two-stage model.} 
The evolutionary process consists of two stages: epidemic spreading and vaccination campaign. 
In the first stage, the population of fraction $x$ have been vaccinated. Then the disease spreads based on the SIR model.
In the second stage, When the disease stops spreading, individuals perform social relationship adjustments and update strategies.
Social link adjustment is incorporated in the vaccination campaign stage,
whereas previous studies concentrate on the strategy updates only in the second stage.}
	\label{figure1}
\end{figure}

We assume that the model consists of two stages \cite{ref11,ref39,ref41,ref42}: epidemic spreading and vaccination campaign which includes social relationship adjustments and strategy updates (see Fig. \eqref{figure1}).
The first stage: epidemic spreads after everybody decides whether to vaccinate or not to according to the last season's vaccination campaign.
Then non-vaccinator will get infected with a certain probability.
The second stage: When the disease stops spreading, individuals adjust their social relationship and update their strategies simultaneously, depending on social bias and the difference in payoff between them and others \cite{ref13}.

The non-vaccinators are likely to be infected with a certain probability,
not only by their neighbors, but also by the people with whom they contact on the road. 	
Thus we assume that the epidemic spreads in a well mixed population \cite{ref31,ref32},
where every individual interacts with all the rest individuals with equal probability.
For social relationship adjustments and strategy updates,
individuals are mainly affected by thy neighbors.
So we assume the social relationship adjustments and strategy updates occur on the network \cite{ref5,ref13},
which is about to vary from time to time.

Every vaccinated individual takes a cost $V> 0$.
This cost $V$ refers to the time and money spent on taking vaccination and its potential side effects such as fever.
We assume that the vaccination is perfect, i.e., whoever takes the vaccination would be free from disease \cite{ref10,ref11}.
For the unvaccinated individuals, some are healthy that pay neither the cost of vaccination nor the cost of treatment. In other words, they are of payoff $0$.
The others are infected with cost $C> 0$.
This cost $C$ includes time and expenses for treatment.
Without loss of generality, we assume that  $C> V>0$.
However, individual's perceived payoff can be different from their actual payoff,
due to underestimating or overestimating the risk of disease \cite{ref15,ref58}.
In order to simplify the model, we take no account of the gap between perceived and actual payoff, and assume that individuals are of perfect recognition.

Vaccination is typically not available any more provided the vaccination campaign is over.
The unvaccinated individuals can be infected when they are surrounded by infected individuals.
We denote the vaccine uptake level by $x$, and probability of getting infected  by $f(x)$.
In general, $f(x)$ is determined by the underlying epidemic dynamics.

\subsection{\textbf{Epidemic spreading in a well mixed population} }
Here we use a  simple Susceptible-Infected-Recovered (SIR) model to capture the dynamics of epidemic spreading.
We have assumed that all individuals are well mixed during the disease spreading.
Susceptible individuals catch infection with rate $\alpha$ if they contact with another infected individual;
the infected recover with a rate $\gamma$ \cite{ref10}.
Noteworthy, individuals do not get infected as long as they get vaccinated,
since we assume that  the effectiveness of vaccination is $100\%$.
In particular, the SIR model results in an infection probability  (see  \ref{appendixa})
\begin{equation}
f(x)=
\begin{cases}
1-\frac{1}{R_{0}\left ( 1-x \right )} & \text{if}  \ 0\leq x<1-\frac{1}{R_{0}} \\
0&\text{if}  \ x\geq 1-\frac{1}{R_{0}}
\end{cases},
\label{eq:refname1}
\end{equation}
where $R_{0}=\alpha/(\gamma+\mu)$ is the basic reproductive ratio \cite{ref10,ref18}.

\subsection{\textbf{Campaign: social relationship adjustments and strategy updates}	}
When diseases stops spreading, there comes the vaccination campaign of the next year. Herein
individuals either adjust their social ties (with probability $\omega$) or adjust their strategies.
We assume a network model,
where nodes represent individuals and links refer to the social ties between individuals.

\subsubsection{\textbf{Social relationship adjustments} }
In the stage of vaccination campaign,
in contrast with previous models,
we assume that network is evolving across time.
This dynamical network arises from social bias \cite{ref13}.
For example:
i) every vaccinator has a high probability of continuing to vaccinate for the next influenza season;
ii) non-vaccinators would like to attach to vaccinators to acquire benefits from potential herd immunity in the next influenza season;
iii) infected non-vaccinators are more likely to be excluded than others.
Typically, during the vaccination compaign,
individuals make up their decisions to vaccinate or not to only once,
whereas they could adjust their social ties many times.
Therefore, we assume that
strategy updates occur rarely compared to link rewirings.

Here we denote the Vaccinated by Strategy $V$,
the Unvaccinated \& Healthy  by Strategy $UH$,
the Unvaccinated \& Infected by Strategy $UI$.
There are three types of individuals,
hence a link $ij$ can be one of the six types ($VV, UHUH, UIUI, VUH, VUI, UHUI$).
We assume that each link breaks with probability $ k_{ij}$
(i.e., probability of link breaking between an individual taking strategy $i$ and an individual taking strategy $j$).

\begin{figure}[!ht]
	\centering
	\includegraphics[width=1\linewidth]{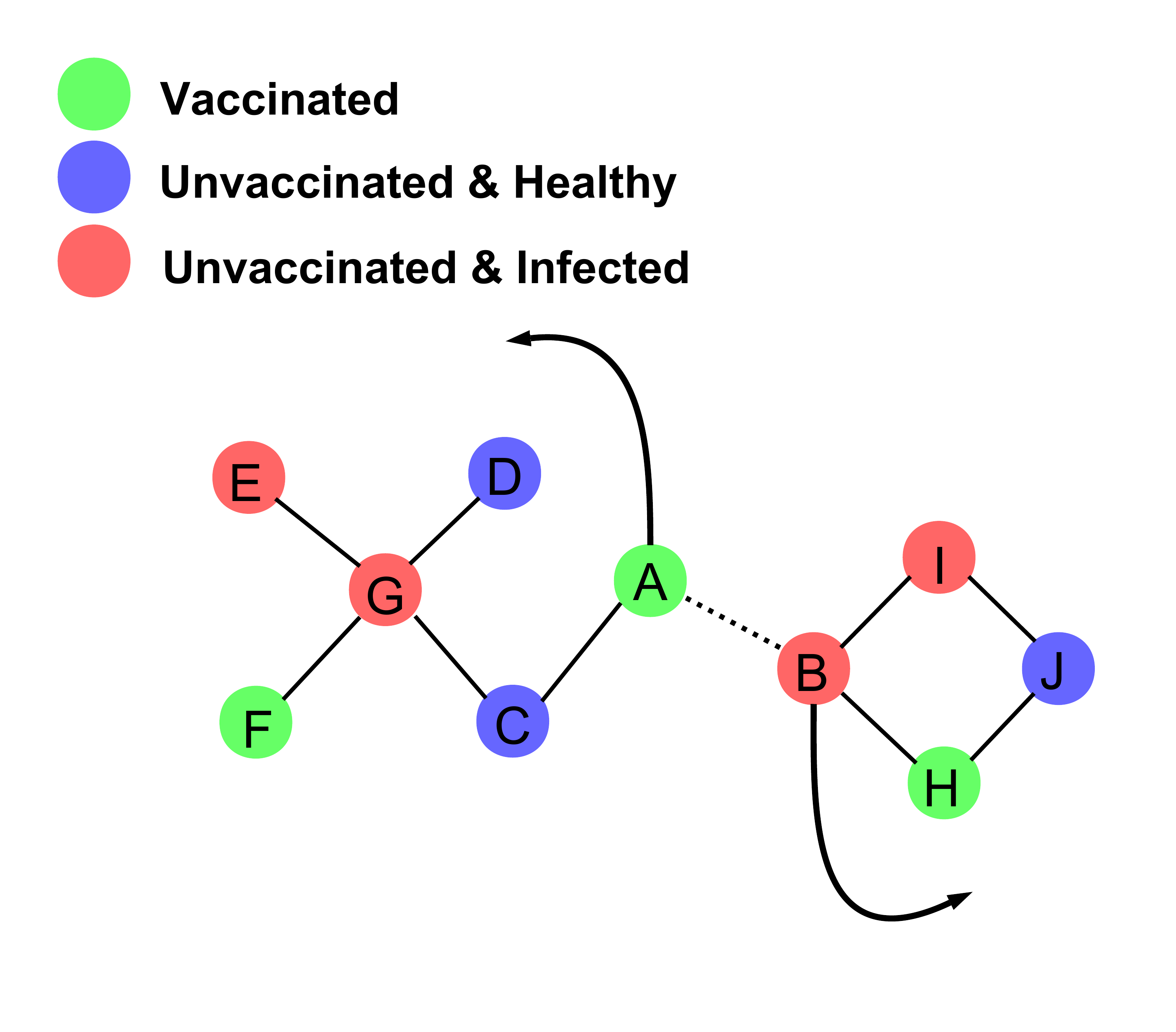}
	\caption{\textbf{The adjustment of social relationships.}
		When the dashed link is selected during the link adjustments,
		it breaks with probability $k_{VUI}$.
		If the link is broken, individual A taking strategy $V$ or individual B taking strategy $UI$ will be selected randomly.
        The selected individual then establishes a new link.
		For example, if individual A is selected,
		it will randomly switch to a neighbor who is not his or her current neighbors (B, D, E, F, G, H, I, J).
		If individual B is selected, it will also randomly switch to a neighbor who is not his or her current neighbors (A, C, D, E, F, G, J). }
	\label{figure2}
\end{figure}

Each time, a link $ij$,
whose two ending nodes are individuals taking strategy $i$ and $j$, is selected.
It breaks with probability $k_{ij}$.
If the link is broken, individual taking strategy $i$ or individual taking strategy $j$ will be entitled to establish a new link with others except his or her current neighbors (see Fig. (\ref{figure2}) for details).

The model captures the dynamical nature of the social network
with the number of links constant over time.
Noteworthy,
social tie adjustments do not happen in the epidemic spreading season.
It does not refer to the fact that susceptible individuals try to break ties with infected neighbors to escape from being infected. 	
Herein social ties are evolving motivated by social bias \cite{ref13}.

\subsubsection{\textbf{Strategy updates}}
For strategy updates, we adopt the imitation rule.
Here, imitation happens via the current network.
Individuals with high payoff are more likely to be followed in strategy \cite{ref33}.
Specifically, we use the Fermi update rule when individual imitates others \cite{ref20,ref21}:
i) A link, namely $ij$, is selected at random;
ii) If the node named individual $i$ at the extremes of the link is selected randomly;
iii) The individual $i$ imitates the strategy of the individual at the other side of the selected link, i.e., $j$, with probability:
\begin{equation}
\frac{1}{1+\exp\left [ -\beta \left ( f_{j}-f_{i} \right ) \right ]},
\label{eq:refname2}
\end{equation}
where $f_{i}$ and $f_{j}$ are the accumulated payoffs of player $i$ and $j$.
And $\beta \geq 0 $ is the selection intensity \cite{ref22}.
The selection intensity indicates how strongly individuals react to payoff gap.
For example, if $\beta$ is large, when $f_{j}-f_{i}>0$,
a small gap between $f_{i}$ and $f_{j}$  will lead to individual $i$ to imitate the strategy of individual $j$  with a great probability.
If $\beta$ is small, i.e., $\beta$ is close to zero,
the probability is close to one half, provided
the difference between $f_{i}$ and $f_{j}$ is not too large.
In other words,
individual $i$ will imitate the strategy of individual $j$ randomly in this case.
Thus the selection intensity also mirrors the rationality of individuals in the population.

\section{Analysis and results}
It is challenging to analytically address the vaccination behavior on dynamical networks.
One of the reasons is that it is not easy to analytically capture the dynamics of networks,
on which imitation takes place.   
In this section,
we make use of the Markov Chain to capture the linking dynamics.
The stationary regime of the linking dynamics can be analytically approximated.
This results in an mean-field equation of the vaccination level with linking dynamics.
The equation facilitates us to 
investigate
how the link rewiring process alters the evolution of vaccination level,
as well as how selection intensity (or rationality) alters the vaccination behavior.

\subsection{\textbf{Vaccination dilemma}}
As mentioned before,
we assume that the population structure is well mixed when the disease spreads.
Naturally, population are divided into three groups.
One is those who takes vaccination.
Because the vaccination is of $100\%$ effective,
these individuals are healthy yet they bear a cost of vaccination $V>0$.
These individuals are of fraction $x$.
For those who do not take vaccination, there are two cases:
one is those who are healthy without any cost.
These unvaccinators benefit from the herd immunity,
so they pay nothing.
The other is those who are infected.
They bear a cost of infection,
including treatment time, expenses and etc, denoted by $C>0$.
An unvaccinated individual gets infected with probability $f(x)$ and the fraction of the unvaccinated individuals is $1-x$.
Hence unvaccinated but healthy individuals is of fraction $(1-x)(1-f(x))$, with cost $0$.
Unvaccinated and infected individuals is of fraction $(1-x)f(x)$, with cost $C>0$ to recover.

We have denoted the Vaccinated by Strategy $V$, the Unvaccinated \& Healthy by $UH$, the Unvaccinated \& Infected by $UI$.
To make it clear, we give the fraction and payoff of the three types of individuals in Table. \eqref{table1}
\begin{table}[ht]
	\centering
	\begin{tabular}{c@{ }|c@{ }|c@{ }|c@{ }}
		\hline
		&Vaccinated&Unvaccinated \& Healthy&Unvaccinated \& Infected\\[6pt]
		\hline
		Notation&$V$& $UH$& $UI$\\[12pt]
		Fraction&$x$& $(1-x)(1-f(x))$& $(1-x)f(x)$\\[12pt]
		Payoff&$-V$& 0& $-C$\\
		\hline
	\end{tabular}
	\caption{The fraction and payoff of all the three types of individuals.}
	\label{table1}
\end{table}

\subsection{\textbf{Strategy updates with linking dynamics: a mean-field analysis}}

When the epidemic season is over, individuals adjust their neighbors motivated by social bias.
When the linking dynamics is much faster than that of the strategy updates, i.e.,
$\omega$ is approaching $1$,
the network topology has already reached its stationary regime
as strategy updates take place \cite{ref5,ref11,ref13,ref23}.

For link rewiring process, the status of a link can be modelled as a random walk in the status space of $\left\lbrace VV, UHUH, UIUI, VUH, VUI, UHUI\right\rbrace $,
i.e., a Markov Chain (see \ref{appendixb}).
The Markov Chain is aperiodic and irreducible,
and thus there exists a unique stationary distribution for each link,
The stationary distribution is given by 	
\begin{equation}
\pi _{ij}=\frac{a(x)(2-\delta _{ij})x_{i}x_{j}}{k_{ij}},
\label{eq:refname3}
\end{equation}
where $i,j\in\{V, UI,VH\}$, $\delta_{ij}$  indicates the Kronecker Delta,
and $a(x)$ is the normalization factor \cite{ref23}.
Noteworthy, $\pi_{ij}$ represents the fraction of $ij$ links in the network when the topology reaches its stationary regime.

For the evolution of the vaccination level $x$, there are four cases:
i) the vaccinated individuals imitate the unvaccinated and healthy individuals;
ii) the unvaccinated and healthy individuals imitate the vaccinated individuals;
iii) the vaccinated individuals imitate the unvaccinated and infected individuals;
iv) the unvaccinated and infected individuals imitate the vaccinated individuals.
Taking the first case as an example, a $VUH$ link is selected with probability $\pi_{VUH}$.
Then the vaccinated individual is selected with probability one half,
this is because individual between the two nodes at the extremes of the link is selected randomly.
And the vaccination individual imitates the strategy of the unvaccinated and healthy individual with probability $[1+\exp\left(\beta \left ( f_{V}-f_{UH} \right ) \right )]^{-1}$.
Similarly, the other three cases are known.
Thus, the dynamics of the vaccination behavior is given by
\begin{equation}
\begin{aligned}
\dot{x} =&-\pi _{VUH}\frac{1}{2}\frac{1}{1+\exp\left [ \beta \left ( f_{V}-f_{UH} \right ) \right ]}\\
&+\pi _{VUH}\frac{1}{2}\frac{1}{1+\exp\left [ \beta \left ( f_{UH}-f_{V} \right ) \right ]}\\
&-\pi _{VUI}\frac{1}{2}\frac{1}{1+\exp\left [ \beta \left ( f_{V}-f_{UI} \right ) \right ]}\\
&+\pi _{VUI}\frac{1}{2}\frac{1}{1+\exp\left [ \beta \left ( f_{UI}-f_{V} \right ) \right ]}
\label{eq:refname4}.
\end{aligned}
\end{equation}

We substitute Eqs. (\ref{eq:refname1})(\ref{eq:refname2})(\ref{eq:refname3}) and Table. (\ref{table1}) into Eq. (\ref{eq:refname4}).
It gives rise to an equation
{\footnotesize
	{\begin{equation}
		\dot{x}=a(x)x\left(1-x\right) \left [  \left({\frac{1}{\displaystyle k_{VUI}}\tanh {\frac{\beta\left(C-V\right)}{2}}}+{\frac{1}{\displaystyle k_{VUH}}\tanh {\frac{\beta V}{2}}}\right)f\left ( x \right )-{\frac{1}{\displaystyle k_{VUH}}\tanh {\frac{\beta V}{2}}}\right ]
		\label{eq:refname18}.
		\end{equation}}}
Because $a(x)$ is positive and does not affect the equilibrium point and its stability,
we get a simplified form of Eq. (\ref{eq:refname18}) by a time rescaling,
{\footnotesize
	{\begin{equation}
		\dot{x}=x\left(1-x\right) \left [  \left({\frac{1}{\displaystyle k_{VUI}}\tanh {\frac{\beta\left(C-V\right)}{2}}}+{\frac{1}{\displaystyle k_{VUH}}\tanh {\frac{\beta V}{2}}}\right)f\left ( x \right )-{\frac{1}{\displaystyle k_{VUH}}\tanh {\frac{\beta V}{2}}}\right ]
		\label{eq:refname5}.
		\end{equation}}}
We are concentrating the limit behavior of the vaccination level,
thus all the analysis are performed based on Eq. \eqref{eq:refname5}.

By setting 	Eq. (\ref{eq:refname5})=0, we find that 
there are two trivial fixed points $0$ and $1$, respectively.
Furthermore,
if 	
\begin{equation}
R_{0}>1+\frac{k_{VUI}}{k_{VUH}}\frac{\tanh {\frac{\beta V}{2}}}{\tanh {\frac{\beta\left(C-V\right)}{2}}}
\label{eq:refname6},
\end{equation}
there exists an internal equilibrium $x^*\in(0,1)$,
which is given by
\begin{equation}
x^{*}=1-\frac{1}{R_{0}} \left(1+\frac{k_{VUI}}{k_{VUH}}\frac{\tanh {\frac{\beta V}{2}}}{\tanh {\frac{\beta\left(C-V\right)}{2}}}  \right) .
\label{eq:refname7}
\end{equation}

Let
$	G(x)= \left({\frac{1}{\displaystyle k_{VUI}}\tanh {\frac{\beta\left(C-V\right)}{2}}}+{\frac{1}{\displaystyle k_{VUH}}\tanh {\frac{\beta V}{2}}}\right)f\left ( x \right )-{\frac{1}{\displaystyle k_{VUH}}\tanh {\frac{\beta V}{2}}}$,
we rewrite Eq. (\ref{eq:refname5}) as
$\dot{x}=x\left(1-x\right)G(x) $.
Since $f(x)$ is a decreasing function,
$G(x)$ is also a decreasing function.
If Eq. (\ref{eq:refname6}) is satisfied, $G(x^{*})$ is always zero and $G(0)$ is always positive.
That is to say, $\dot{x}>0$ when $x<x^{*}$, and $\dot{x}<0$ when $x>x^{*}$.
So when Eq. (\ref{eq:refname6}) is satisfied,
$x^{*}$ is an internal stable equilibrium.
If Eq. \ref{eq:refname6} does not hold,
there is no internal equilibrium with $x^*=0$ the only stable equilibrium.

By analyzing the internal equilibrium $x^{*}$,
we get a main result:
if there is an internal equilibrium,
then it has to be stable.

\subsection{\textbf{Social bias and rationality}}
Based on Eq. (\eqref{eq:refname7}),
we observe that the final vaccination level is determined by 
the social bias, i.e., the linking dynamics, $\frac{k_{VUI}}{k_{VUH}}$ and 
the rationality $\beta$,
besides the payoff entries of the vaccination $C$ and $V$.
We try to figure out
i) how the social bias, i.e. linking dynamics, alters the vaccination level;
and ii) how the rationality alters the vaccination level.
All the analysis are performed for a sufficient large basic reproductive ratio determined by Eq. (\ref{eq:refname6}).

\subsubsection{\textbf{Social bias}}
\begin{figure}[!ht]
	\centering
	\includegraphics[width=1\linewidth]{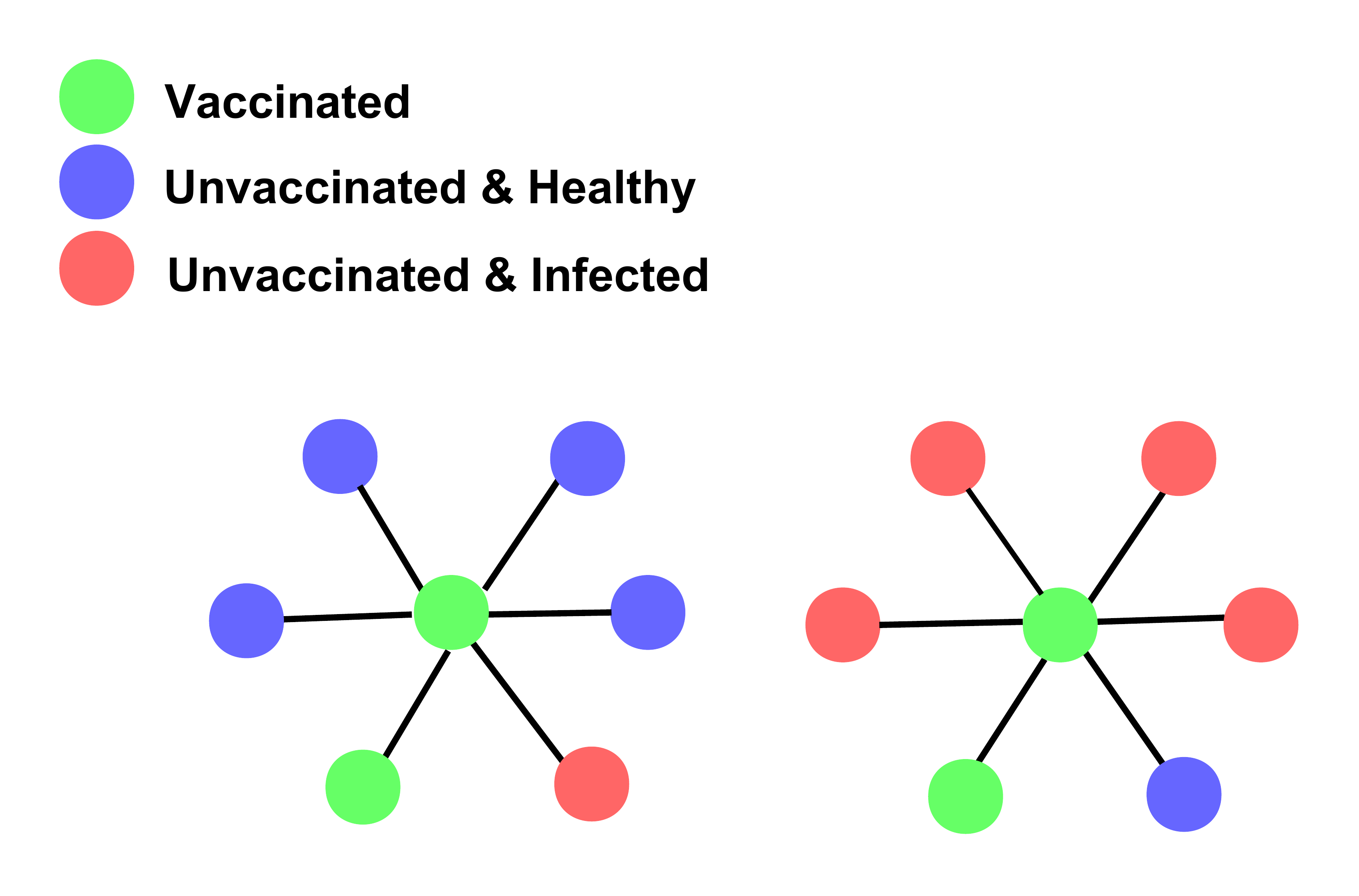}
	\caption{\textbf{Illustration of the network configuration driven by linking dynamics.} 
$\tfrac{k_{VUI}}{k_{VUH}}$ is the breaking probability ratio between vaccinated-infected-unvaccinator and vaccinated-healthy-unvaccinator links.
The smaller the ratio $\tfrac{k_{VUI}}{k_{VUH}}$  is, the more likely a vaccinator meets an infected unvaccinator than meets a healthy unvaccinator.
The left panel shows a local network configuration of large $\tfrac{k_{VUI}}{k_{VUH}}$,
whereas the right panel indicates that of a small $\tfrac{k_{VUI}}{k_{VUH}}$.
The local network configuration similar to the right panel is key for a high vaccination level.}
	\label{figure8}
\end{figure}

From Eq. (\ref{eq:refname7}),
we get that $x^{*} $ is an increasing function of $k_{VUH}$,
but a decreasing function of $k_{VUI}$.
We show this in Fig. \eqref{figure5}.
Intuitively, from the stationary distribution for each link Eq. (\ref{eq:refname2}),
$\pi_{ij}$ is a decreasing function of $k_{ij}$.
In other words, the smaller the $k_{ij}$ is, the more $ij$ links there are.
Or the stronger $ij$ ties are, the larger the number of $ij$ links is.
If $k_{VUH}$ is small, the $\pi_{VUH}$ is large,
i.e., the number of unvaccinated and healthy individuals around the vaccinator  increases.
This has been shown on the left hand side of Fig. \eqref{figure8}.
When strategy updates,
the probability becomes larger that vaccinators select an unvaccinated and healthy individual to imitate.
In addition, the payoff of unvaccinated and healthy individual is larger than vaccinated individual,
so the probability with which vaccinator imitates the strategy of unvaccinated and healthy individual becomes even larger.
Consequently fewer individuals get vaccinated.

If $k_{VUI}$ is small, $\pi_{VUI}$ is large,
i.e., the number of infected unvaccinators around the vaccinator increases,
shown on the right hand side of Fig. \eqref{figure8}.
At the same time, infected individuals are isolated from the other unvaccinated via the vaccinated individual.
When strategy updates,
the likelyhood that infected non-vaccinators select a vaccinated individual becomes larger.
Infected unvaccinators are more likely to imitate the strategy of vaccinators,
since the payoff of infected unvaccinator is smaller than that of the vaccinators'.
Then more people will take vaccination.		

\subsubsection{\textbf{Rationality}}
\label{sec:selectionintensity}
Noteworthy,
Eq. (\ref{eq:refname7}) captures the evolution of vaccination behavior
on a complex dynamical network for any selection intensity.
Taking the internal equilibrium $x^{*}$ as a function of selection intensity, we get
{	\scriptsize
	\begin{equation}
	\begin{aligned}
	\frac{d x^{*} }{d \beta }=-\frac{k_{VUI}}{k_{VUH}}\frac{1}{R_{0}}\left ( {{\frac{V}{2}{\rm sech}\; ^2\left(\frac{V}{2}\beta \right)\,\tanh \left(\frac{C-V}{2}\beta \right)-\frac{C-V}{2}\tanh \left(\frac{V}{2}\beta \right)\,{\rm sech}\; ^2\left(\frac{C-V}{2}\beta\right)}\over{ \tanh ^2\left(\frac{C-V}{2}\beta \right)}} \right )
	\label{eq:refname8}.
	\end{aligned}
	\end{equation}
}
When the ratio of $C$ to $V$ exceeds $2$,  $\frac{dx^{*}}{d \beta }$
is always negative.
Hence, $x^{*} $ is a decreasing function of $\beta$.
In other words, if $C>2V$, the stronger the reaction of individuals to the payoff gap is,
i.e., the stronger the selection intensity is,
the less people get vaccinated when the disease breaks out.
When the ratio of $C$ to $V$ is below 2,
the derivative of the equilibrium, i.e., $\frac{dx^{*}}{d \beta }$, is always positive.
Hence, $x^{*} $ is an increasing function of $\beta$.
In other words, if $C<2V$, the stronger the reaction of individuals to the payoff gap is,
the more people get vaccinated when the disease breaks out. (see \ref{appendixc} for more details)

It has been suggested that weak selection promotes the spread of altruism and strong selection impedes altruism \cite{ref25}.
When the ratio of $C$ to $V$ is beyond $2$ and the selection intensity is weak, i.e., the medical cost is high,
people are more likely to take vaccination to avoid the expensive medical cost,
which improves the herd immunity and protects the unvaccinated connected to them at the same time.
Here, vaccination can be seen an altruism behavior,
because society benefits from herd immunity which is contributed by vaccinated individuals \cite{ref2,ref35,ref36}.
When the selection intensity is strong,
people are sensitive to the payoff gap.
At the same time, the fraction of people around them getting vaccinated is large,
which leads to a high herd immunity.
The probability of successful hitchhiking without paying anything is high \cite{ref2,ref3,ref4},
so individuals prefer not to vaccinate and try to benefit from herd immunity to keep healthy.

When the ratio of $C$ to $V$ is below 2, i.e., the medical cost is low.
People are more likely not to vaccinate to stay away from the side effects of the vaccine when the selection intensity is weak.
When the selection intensity is strong,
they find it is better to get vaccinated.
Because the herd immunity is low, which makes it easily get infected \cite{ref2,ref3,ref4}.
Since the medical cost $C$ is always larger than the vaccination cost $V$,
people are more likely to get vaccinated when the selection intensity is strong.

When the selection intensity $\beta$ is quite strong,
both $\tanh {\frac{\beta\left(V\right)}{2}}$ and $\tanh {\frac{\beta\left(C-V\right)}{2}}$ approach  $1$ based on Eq. \eqref{eq:refname7}.
In this case, the vaccination level $x^{*}$ approaches
\begin{align}
1-\frac{1}{R_{0}} \left(1+\frac{k_{VUI}}{k_{VUH}}\right).
\label{strongSEL}
\end{align}
On the one hand, it implies that the final vaccination level is independent of neither $C$ nor $V$ (see Fig. \eqref{figure4}).
This mirrors the reality:
Under strong selection,
the infected individual will imitate vaccinated individual with probability approaching 1.
Similarly, the vaccinated individual will imitate the infected individual with probability approaching 0.
So the vaccination level is no longer affected by $C$ or $V$,
provided the selection intensity $\beta$ is strong.
The final vaccination level is sensitive to the ratio $C/V$,
only if the selection intensity $\beta$ is weak.
On the other hand,
the final vaccination level is determined by both the basic reproductive ratio $R_0$
and the link breaking probability ratio between $V-UI$ and $V-UH$.
The larger the basic reproductive ratio $R_0$ is,
the higher the final vaccination level reaches,
which is consistent with previous works \cite{ref5,ref43}.
The ratio between $k_{VUI}$ and $k_{VUH}$ mirrors the social bias in the population.
When the ratio is one.
no social bias is present, as if in a well-mixed population.
And the final vaccination level $1-\tfrac{2}{R_0}$ degenerates to that in a well mixed population.
If the ratio $\tfrac{k_{VUI}}{k_{VUH}}$ is smaller, i.e., vaccinated individuals are likely to connect with the infected unvaccinators.
The final vaccination level increases.
This is because,
infected individuals are isolated from the other unvaccinated via the vaccinated individual.
It results in a population configuration,
where unsuccessful free-riders are more likely to connect with vaccinators.
Considering the population structure is for social imitation,
unsuccessful free-riders are likely to imitate other vaccinators than in a well mixed population.
Therefore, the vaccination level increases.
In fact the final vaccination level in a dynamical network under strong selection, i.e., Eq. \eqref{strongSEL}
implies that social bias can be interpreted as a reproductive ratio rescaling in a well-mixed population.
The larger the ratio $\frac{k_{VUI}}{k_{VUH}}$ on a dynamical network is,
i.e., the greater social bias is,
the smaller the effective basic reproductive ratio in the well mixed population is.
This results in a low vaccination level.
Following the same argument,
we have that the vaccination level increases, if the ratio $\tfrac{k_{VUI}}{k_{VUH}}$ is small.
Furthermore, the maximum vaccination level here is given by $1-\tfrac{1}{R_0}$,
which is obtained when the vaccinator-infected-unvaccinated ties are much more close than  vaccinator-healthy-unvaccinated ones.
Noteworthy, this vaccination level is higher than that in the well-mixed population,
given by $1-\tfrac{2}{R_0}$.
thus the social bias in imitation would drive the population  away from the Nash equilibrium to the social optimum vaccination level. 

\section{Agent-based simulation}
In this section, we perform an agent-based simulation  to validate theoretical analysis.
Simulation procedures are as follows.
\begin{enumerate}
\item Initially, $N$ individuals are situated on the vertices of a network.
Each individual has exactly $k$ neighbors. Randomly, $Nx$ individuals choose to take vaccination, while others not, where $x \in [0,1]$ is initial vaccination level. Thus yields infection probability $f(x)$ for unvaccinated individuals. That is to say, $N(1-x)f(x)$ unvaccinated individuals get infected at random.
\item In each generation, we generate a random real number $r \in (0,1)$. If $r>\omega$, we perform one strategy update accompanying an epidemic season(\textbf{Go to step. (3)}). Otherwise, we execute social relationship adjustment(\textbf{Go to step. (4)}).
\item If a strategy update occurs, then a link is selected at random whose ending nodes are namely Alice and Bob.
Alice or Bob is selected randomly to update his or her strategy. Without loss of generality let us assume that Bob is to update his strategy. His payoff and Alice's are denoted by $p_\text{Bob}$ and $p_\text{Alice}$ respectively. With probability
\begin{align*}
\frac{1}{1+exp[-\beta(p_\text{Alice}-p_\text{Bob})]},
\end{align*}
Bob takes Alice's strategy. Otherwise, Bob keeps his strategy.
Then there comes a new epidemic season.
As vaccination level $x$ alters in the process of strategy update, we have a new infection rate $f(x)$. Similarly, $N(1-x)f(x)$ unvaccinated individuals get infected randomly. It is noteworthy that all the previous infected individuals recovered at the end of last epidemic season. Hence infected individuals in current epidemic season are independent of previous ones.\\
\textbf{Go back to step. (2)}.
\item If a social relationship adjustment occurs, then a link is selected at random. The two nodes at the two extremes of the selected link take strategies $i$ and $j$ respectively. With probability $k_{ij}$, the link breaks. If the link breaks, then $i$ or $j$ is entitled to establish a new link with anyone who is not his or her current neighbors. Otherwise, the link remains connected. If the operations above result in any isolated space(no neighbors), then we countermand all the operations and perform social relationship adjustment again.\\
\textbf{Go back to Step. (2)}.
\end{enumerate}

\begin{figure}
	\centering
	\includegraphics[scale=0.6]{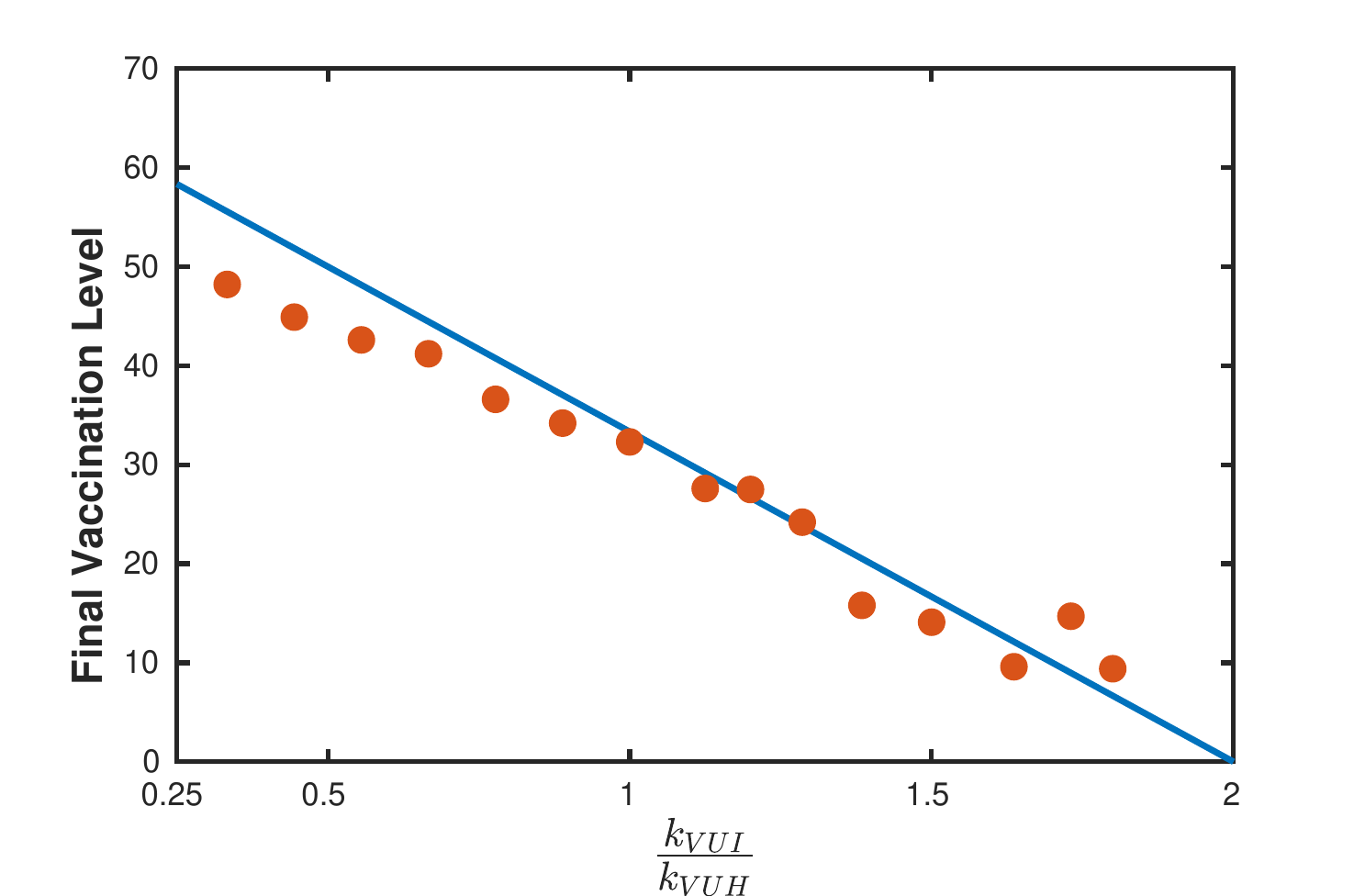}
	\caption{
		\textbf{The vaccination level as a function of $\frac{k_{VUI}}{k_{VUH}}$.} The line indicates numerical solutions based on Eq. \eqref{eq:refname7}. As $k_{VUI}$ increases or $k_{VUH}$ decreases, i.e., $\frac{k_{VUI}}{k_{VUH}}$ goes up, the vaccination level goes down. The points represent the agent-based simulation which agree with the numerical simulation. Each data point is the average of 30 independent runs with initial vaccination level of 50\%. And in each run, the final vaccination level is estimated  with the largest sojourn time over $10^7$ generations. The final vaccination level is theoretically captured by the stable equilibrium of the differential equation Eq. (\eqref{eq:refname5}). 
In addition, $\beta=10, 1-\omega=0.001, k_{VV}=k_{UHUH}=k_{UIUI}=k_{UHUI}=0.5, N=100, k=4$ for all data points and curve above.}
	\label{figure5}
\end{figure}

\begin{figure}
	\centering
	\includegraphics[scale=0.6]{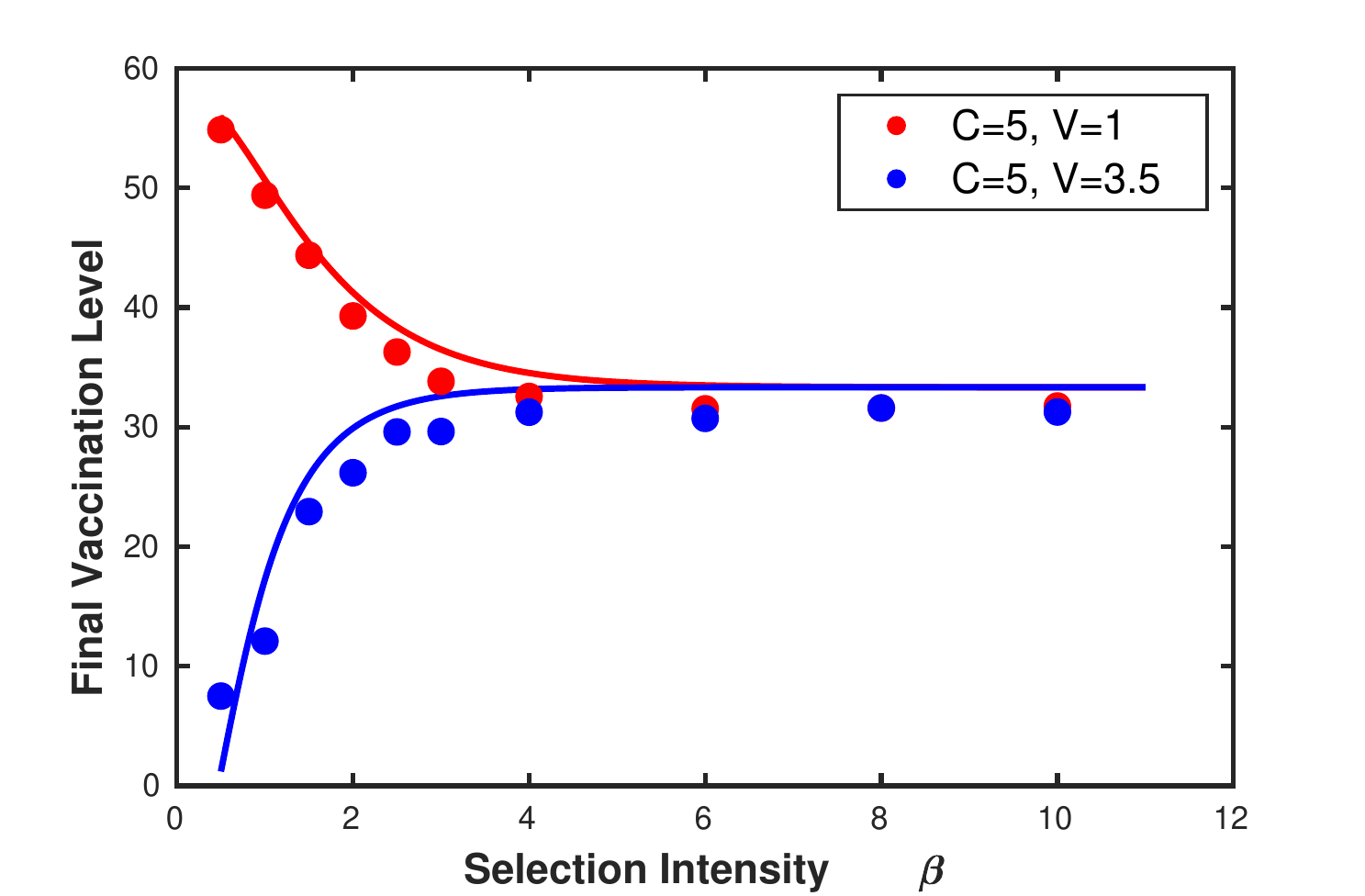}
	\caption{
		\textbf{The vaccination level as a function of rationality characterized by selection intensity $\beta$.} The curves indicate numerical solutions based on Eq. \ref{eq:refname7}. When $C<2V$, the vaccination level goes up as $\beta$ increases. On the contrary when $C>2V$, the fraction of vaccination goes down as $\beta$ increases. The above two cases tend to agree each other as long as $\beta$ large enough. That is to say, when $\beta$ is too large, the relation between $C$ and $2V$ has no influence on the vaccination level. The points represent the agent-based simulation which basically agree with the numerical simulation. Each data point is the average of $30$ independent runs with initial vaccination level of 50\%. And in each run, we set the state with largest sojourn time over $10^7$ generations as the equilibrium state. 
In addition, $1-\omega=0.001, k_{VV}=k_{UHUH}=k_{UIUI}=k_{UHUI}=0.5, k_{VUI}=k_{VUH}=0.8, N=100, k=4$ for all the curves and data points above.}
		\label{figure4}
\end{figure}

    In Fig. \eqref{figure5}\eqref{figure4},
    there is a deviation between theoretical and simulated values, which is due to the stochasticity in link rewiring.
    The vaccination level is predicted by deterministic equations.
    We assume that when strategy updates, the adjustment of social relationships has already reached a steady state.
    However, in the simulation, the strategy can update when link rewiring have not reached the steady state.

\section{Conclusion and discussion}
Preemptive vaccination has been the principal strategy to control influenza \cite{ref26,ref27}.
Vaccination is similar to multi-player snowdrift game:
non-vaccinators try to exploit the herd immunity created by the vaccinators,
whereas vaccinators pay costs to get the immunity.
Those who do not shovel in the snow drift would benefit from those who shovel,
whereas those who shovel pay costs themselves to go home.
Here, vaccination can be seen an altruism behavior,
because non-vaccinators benefit from herd immunity which is contributed by vaccinators \cite{ref2,ref35,ref36}.
Individuals in vaccination dilemma always tend to do what others do not do,
as individuals in snow drift game do.
However, they are not the same: in snowdrift game, if individuals choose to defect,
they will definitely benefit, provided there are enough cooperators \cite{ref43}.
In other words, as long as the minimum number of people required to clear snow is exceeded, those who do not shovel will go home paying nothing. For vaccination, however, unvaccinated individuals can still be infected,
even if there is herd immunity.
The risk of being infected is a double-sword in the vaccination dilemma.
One the one hand, unsuccessful free riders, i.e., the infected non-vaccinators gain less than the vaccinated,
and it gives rise to an increase in vaccination level;
On the other hand, vaccinators gain less than successful free riders, i.e., healthy unvaccinated individuals,
and it yields that vaccination level would decrease.
It is an arm race between these two tendency in a dilemma of vaccination via imitation.

Here, we take into account of the dynamical nature of the social network on which imitation takes place.
In contrast with previous works \cite{ref5,ref11,ref30,ref38,ref39,ref40},
we combine three dynamics together:  i) epidemics dynamics  ii) linking dynamics, and iii) imitation dynamics.
The three dynamics feedback with each other to adjust the vaccination level:
When disease breaks out, the probability of an unvaccinated individual getting infected is determined by the vaccination level.
When the disease is over, individuals adjust their ties via social bias or preferential attachment \cite{ref13}.
The resulting network configuration determines 'who imitates whom',
which is crucial for imitation dynamics.
For example, infected non-vaccinators are more likely to be excluded than others.
They adjust their social relationships and update strategies to get potential herd immunity.
It is found that it is helpful to inhibit the outbreak of disease,
when the symmetry between epidemic and information transmission network breaks \cite{ref30}.
Therefore, it is often that  epidemic spreading and vaccination campaign are assumed to occur in two different networks as in \cite{ref43,ref45}.
We inherit this assumption.
In particular, we notice that
non-vaccinators are likely to be infected by anyone who has contacts with them.
But during vaccination campaign, they only imitate their neighbors' strategies via the social network.
Therefore we assume that the epidemic spreads in a well mixed population \cite{ref31,ref32} and the vaccine campaign evolves on a stochastic network, which is more practical.
What's more, we assume that the SIR model has already reached a steady state when vaccination campaign takes place.
For vaccination campaign, we assume that strategy updates happen rarely compared to social relationship adjustments.

Whether or not to vaccinate can be seen as whether or not to cooperate with others.
We describe the social relationship adjustments through linking dynamics,
which is often used to study the relationship between cooperators and defectors \cite{ref23,ref50,ref55,ref56,ref57}.
For strategy updates, imitation process typically assumes that an individual is selected randomly and then choose a neighbor around the  selected individual randomly to imitate the its strategy \cite{ref5,ref23,ref11,ref30,ref31,ref38}.
Noteworthy, we describe strategy updates in terms of selecting a link first and then choosing an individual between two nodes at the extremes of the link randomly to imitate the other one.
This method results in a simple deterministic equation Eq. \eqref{eq:refname4},
which is the key for theoretical analysis.

We find that after disease spreading, reducing the interaction between vaccinated individuals and hitchhiking individuals, or increasing interaction between vaccinated individuals and infected individuals, are helpful to control the disease.
Intuitively, reducing the interaction between vaccinated individuals and successful hitchhiking individuals decreases the chance that
these two types individuals meet.
Thus vaccinators are less likely to meet the successful free-riders,
who is better off than vaccinators.
Further, It results in a low probability a vaccinator taking no vaccination next year.
Thus it promotes vaccination.
Similarly, increasing interaction between vaccinated individuals and infected individuals yields a network configuration
where infected non-vaccinators meet more often the vaccinators.
In this case, vaccinators are better off than infected non-vaccinators.
Thus non-vaccinators are more likely to take vaccination via imitation next year.
In fact,
there are two kinds of defectors in the vaccination dilemma, the lucky ones (healthy) and the unlucky ones (infected).
Thus there are two kinds of cooperator-defector links.
Our results show that the fragility of the two kinds of cooperator-defector play the opposite role in promoting vaccination level.
This arm race arises from the double-sword effect mentioned in the first paragraph in the discussion.
Furthermore, it is not valid any more that
the more fragile the relationships between cooperators and defectors are, the higher cooperation level is achieved \cite{ref50},
which is a well-known result in cooperation on dynamical networks.

In addition, we study how the vaccination level alters as a function of selection intensity (or rationality),
which is typically challenging in evolutionary game theory \cite{ref5,ref11,ref42}.
We make use of the deterministic replicator equation to overcome this obstacle.
We find that the relative cost, i.e.,  the medical cost over the vaccine cost, plays a key role:
When the relative cost is beyond $2$, the weaker the selection intensity is,
the more individuals take vaccination to avoid the expensive medical cost and improve the herd immunity at the same time.
When the relative cost is below $2$, the weaker the selection intensity is, the fewer individuals take vaccination.
This does not go along with previous results that weak selection promotes the spread of altruism and strong selection impedes altruism \cite{ref25}.
When the selection intensity is sufficiently strong, the relative cost hardly affects the vaccination level.
This is because people are oversensitive to the payoff gap.
A small payoff gap will make them imitate others' strategies with probability approaching $1$.
And in particular under strong selection,
we find that the vaccination dilemma on an evolving network mirrors the vaccination dilemma in the well-mixed population with a rescaled basic reproductive ratio.
It indicates that proper social bias in imitation could be equivalent with reducing the basic reproductive ratio.
It implies the social reconnection or adjustment,
even when the epidemics season is over,
can still play a key role in vaccination taking behavior.
Yet noteworthy, the social bias here exists in the vaccination campaign rather than in the epidemic season.
It has nothing to do with the isolation in the epidemic season.

Let us further consider that robustness of the results under different ways of payoff collection.
For instance,  the opponents' payoff can be calculated as the average payoff who take the same strategy as his opponents \cite{ref41,ref42}. We find that strategy update rules are equivalent to strengthen or dilute selection intensity. For example, in \cite{ref41}, they compared two strategy update rules. One is an original rule via which an individual just compares his payoff with the payoff of a randomly selected neighbor.
The other considers a new rule where an individual compares his payoff with the average payoff who take the same strategy as his opponent.
We find the new rule mirrors diluting the selection intensity, compared to the original rule (see \ref{appendixd}).
In other words, we could rescale selection intensity to investigate this issue based on Section \ref{sec:selectionintensity}.

To sum up,
we incorporate the dynamical nature of social network into vaccination.
And we find that it enhances the vaccination level,
if infected non-vaccinated individuals meet more often the vaccinators,
or if the healthy non-vaccinators interact less often with the vaccinators.
Furthermore, we find that the dynamical nature of social network would be equivalent with rescaling basic reproductive ratio.

\begin{appendix}
\section{Infection probability of the unvaccinated}
\label{appendixa}
We adopt SIR model to capture dynamics of epidemic spreading \cite{ref10}.
There are three types of individuals:
 susceptible individuals (S),
 infected individuals (I),
 recovered individuals (R).
The dynamics of the SIR is given by
	
	\begin{equation}
	\frac{dS }{dt }=\mu \left(1-x \right) -\alpha SI-\mu S
	\label{eq:refname13},
	\end{equation}
	
	\begin{equation}
	\frac{dI }{dt }=\alpha SI -\gamma I-\mu I
	\label{eq:refname14},
	\end{equation}
	
	\begin{equation}
	\frac{dR }{dt }=\mu x +\gamma I -\mu R
	\label{eq:refname15},
	\end{equation}
	Here, $\mu$ is the birth and death rate.
	Since we assume the population is constant in size,
	$\alpha$ is the mean transmission rate,
	$\gamma$ is the mean recovery rate,
	$x$ is the vaccination level.
	$R_{0}$ is the basic reproductive ratio.
	From the Eq. \eqref{eq:refname14}, we get $R_{0}=\alpha/(\gamma+\mu)$.
	If $R_{0}\le1$ and $dI/dt$ is negative, the disease cannot persist.
	When the the proportion of the population $(S,I,R)$ is stable,
	we get $S^{*}=1/R_{0}, I^{*}=\mu[R_{0}(1-x)-1]/\alpha, R^{*}=1-I^{*}-S^{*}$.
	By setting $I^{*}>0$,
	we find the disease cannot be eradicated, when the vaccination level $x<1-1/R_{0}$.
	When the vaccination level $x\ge1-1/R_{0}$,
	people will benefit from the herd immunity and the disease will be eradicated.
	So when $x\ge1-1/R_{0}$, $f(x)=0$, i.e., the disease will be eradicated.
	
	The probability for each unvaccinated individual, when the vaccination uptake level is $x$, being infected is denoted by $f(x)$.
	The death rate is  $\mu$
	and the infection rate is $\alpha I^{*}$  of an unvaccinated individual.
	In addition, the death and infection are two independent  Poisson processes,
    hence $f(x)$ is denoted by $\alpha I^{*}/(\alpha I^{*}+\mu)$.
	We arrive at
	\begin{equation}
	f(x)=
	\begin{cases}
	1-\frac{1}{R_{0}\left ( 1-x \right )} & \text{if}  \ 0\leq x<1-\frac{1}{R_{0}} \\
	0&\text{if}  \ x\geq 1-\frac{1}{R_{0}}
	\end{cases}.
	\label{eq:refname16}
	\end{equation}
	
\section{The stationary regime of the dynamical network}
\label{appendixb}
Every individual has three strategies to choose from $\left\lbrace V,UH,UI\right\rbrace $, hence there are six types of links $ij$ ($ij\in\left\lbrace VV, UHUH, UIUI, VUH, VUI, UHUI\right\rbrace $).
	Each link is broken with probability $ k_{ij}$
	(i.e., probability of link breaking between individual taking strategy $i$ and individual taking strategy $j$).
	We select a link named $m$.
	If the link does not break, $ m^{t+1}=m^{t}$.
	If the link breaks, a new link is introduced.
	The new link is only related to $m^{t}$. [To Prof. Wu, what we wanna say is Markov Property, once $m^{t}$ is known, $m^{t+1}$ is independent of $m^{t_0}$ where $t_0<t$]
	In other words,
	the status of next moment is only determined by the former moment,
	which is the property of Markov.
	A Markov Chain with the transition matrix $V_{(AB)(CD)}$ can be constructed to capture the link dynamics.
	The $V_{(AB)(CD)}$ indicates the transition probability
	that a $AB$ link  turns to a $CD$ link in one time step.
    Therefore the conditional transition matrix $V$ is given by
	\begin{equation}
	\tiny{
		\begin{smallmatrix}
		\bordermatrix{%
			& VV   & UHUH   &UIUI  &VUH   &VUI   &UHUI  \cr
			VV    &1-k_{VV}(1-x_{V}) &0 &0 &k_{VV}x_{UH}  &k_{VV}x_{UI} &0 \cr
			UHUH  &0 &1-k_{UHUH}(1-x_{UH})  &0 &k_{UHUH}x_{V} &0 &k_{UHUH}x_{UI} \cr
			UIUI  &0 &0 &1-k_{UIUI}(1-x_{UI}) &0 &k_{UIUI}x_{V}  &k_{UIUI}x_{UH} \cr
			VUH   &\frac{k_{VUH} x_{V}}{2} &\frac{k_{VUH} x_{UH}}{2} &0 &1-\frac{k_{VUH}(1+ x_{UI})}{2} &\frac{k_{VUH} x_{UI}}{2} &\frac{k_{VUH} x_{UI}}{2}\cr	
			VUI   &\frac{k_{VUI} x_{V}}{2} &0 &\frac{k_{VUI} x_{UI}}{2} &\frac{k_{VUI} x_{UH}}{2} &1-\frac{k_{VUI}(1+x_{UH})}{2} &\frac{k_{VUI}x_{UH}}{2} \cr   	
			UHUI  &0 &\frac{k_{UHUI} x_{UH}}{2} &\frac{k_{UHUI} x_{UI}}{2} &\frac{k_{UHUI} x_{V}}{2} &\frac{k_{UHUI} x_{V}}{2} &1-\frac{k_{UHUI}(1+ x_{V})}{2}
		}
		\end{smallmatrix}}
	\end{equation}.
	\label{Eq:matrix1}
	
	Note that the Markov chain is aperiodic and irreducible,
	when $x_{V}x_{UH}x_{UI} \prod_{ij} k_{ij}\ne0$.
	Hence, there exists a unique stationary distribution $\pi=(\pi_{VV},\pi_{UHUH},\pi_{UIUI},\pi_{VUH},\pi_{VUI},\pi_{UHUI})$.
	The stationary distribution $\pi$ is determined by the equation $\pi V=\pi$ \cite{ref23}.
	We find that
	\begin{equation}
	\pi _{ij}=\frac{a(x)(2-\delta _{ij})x_{i}x_{j}}{k_{ij}}
	\label{eq:refname17},
	\end{equation}
	where $\delta_{ij}$  indicates the Kronecker Delta.
	Here $a(x)$ is the normalization factor, where $x=( x_{V}, x_{UI}, x_{UH})$, is given by $a(x)=[(x_{V}^{2}/k_{VV})+(x_{UH}^{2}/k_{UHUH})+(x_{UI}^{2}/k_{UIUI})+(2x_{V}x_{UH}/k_{VUH})+(2x_{V}x_{UI}/k_{VUI})+(2x_{UH}x_{UI}/k_{UHUI})]^{-1}$.
	
\section{Vaccination level  as a function of the selection intensity}
\label{appendixc}
	In the section \ref{sec:selectionintensity}, we have get the derivative of the equilibrium as a function of the selection intensity $\frac{\partial x^{*} }{\partial \beta }$.
	
	We set
	{\footnotesize
		\begin{equation}
		\begin{aligned}
		H(\beta)={\frac{V}{2}{\rm sech}\; ^2\left(\frac{V}{2}\beta \right)\,\tanh \left(\frac{C-V}{2}\beta \right)-\frac{C-V}{2}\tanh \left(\frac{V}{2}\beta \right)\,{\rm sech}\; ^2\left(\frac{C-V}{2}\beta\right)}
		\label{eq:refname23},
		\end{aligned}
		\end{equation}}
	then we get
		\begin{equation}
		\frac{\partial x^{*} }{\partial \beta }=-\frac{k_{VUI}}{k_{VUH}}\frac{1}{R_{0}}{{H(\beta)}\over{ \tanh ^2\left(\frac{C-V}{2}\beta \right)}}.
		\end{equation}
		
	The derivative of $H(\beta)$ as a function of selection intensity $\beta$ is
	{\footnotesize
		\begin{equation}
		\frac{\partial H(\beta) }{\partial \beta}=2{\left( {\left( \frac{C-V}{2}\right)\; ^2 {\rm sech}\; ^2\left(\frac{C-V}{2}\beta \right)\,-\left( \frac{V}{2}\right)\; ^2 {\rm sech}\; ^2\left(\frac{V}{2}\beta \right)\,}\right) \tanh \left(\frac{C-V}{2}\beta\, \right)\tanh \left(\frac{V}{2}\beta \right)}.
		\end{equation}}
		
	Let's denote
	{\footnotesize
		\begin{equation}
		\begin{aligned}
		T(\beta)= \frac{C-V}{2} {\rm sech}\left( \frac{C-V}{2}\beta \right) \,- \frac{V}{2}{\rm sech}\left(\frac{V}{2}\beta \right)
		\label{eq:refname9}.
		\end{aligned}
		\end{equation}}
	We have that
	{	\footnotesize
		
		\begin{equation}
		\begin{aligned}
		\frac{\partial H(\beta) }{\partial \beta}=2{T(\beta)\left( \left( \frac{C-V}{2}\right) {\rm sech}\left(\frac{C-V}{2}\beta \right)\,+\left( \frac{V}{2}\right){\rm sech}\left(\frac{V}{2}\beta \right)\right) \tanh \left(\frac{C-V}{2}\beta\, \right) \tanh \left(\frac{V}{2}\beta \right)}
		\label{eq:refname10}.
		\end{aligned}
		\end{equation}}
	In order to investigate the sign of $T(\beta)$,
	let's first analyze  the relationship between $\frac{C-V}{V}$ and $\frac{\rm sech\left(\frac{V}{2}\beta \right)}{\rm sech\left(\frac{C-V}{2}\beta \right)}$.
	{	
		\begin{equation}
		\begin{aligned}
		\frac{\partial \frac{\rm sech\left(\frac{V}{2}\beta \right)}{\rm sech\left(\frac{C-V}{2}\beta \right)}}{\partial \beta} = {{
				\left( \frac{C-V}{2}\tanh \left(\frac{C-V}{2}\,\beta\right)-
				\frac{V}{2}\, \tanh \left( \frac{V}{2}\,\beta\right) \right)\,{\rm sech}\; \left(\frac{V}{2}\,\beta\right)}\over{ {\rm sech}\; \left( \frac{C-V}{2}\,\beta\right)}}
		\label{eq:refname11}.
		\end{aligned}
		\end{equation}}
	
	If $\frac{C-V}{2}>\frac{V}{2}>0$, i.e., $\frac{C-V}{V}>1$,
	we obtain that  $\lim \limits_{\beta \to +\infty} \frac{\rm sech\left(\frac{V}{2}\beta \right)}{\rm sech\left(\frac{C-V}{2}\beta \right)}=\lim \limits_{\beta \to +\infty} \frac{\exp(\frac{C-V}{2}\beta)}{\exp(\frac{V}{2}\beta)}=+\infty>\frac{C-V}{V}$
	and $\frac{\partial \frac{\rm sech\left(\frac{V}{2}\beta \right)}{\rm sech\left(\frac{C-V}{2}\beta \right)}}{\partial \beta}>0$.
	This implies that
	$\frac{\rm sech\left(\frac{V}{2}\beta \right)}{\rm sech\left(\frac{C-V}{2}\beta \right)}$
	is an increasing function of $\beta$.
	On the other hand,
	$\lim \limits_{\beta \to 0} \frac{\rm sech\left(\frac{V}{2}\beta \right)}{\rm sech\left(\frac{C-V}{2}\beta \right)}=1<\frac{C-V}{V}$,
	so there exists $\beta^{*}>0$ which makes $\left( \frac{C-V}{2}\right) {\rm sech}\left(\frac{C-V}{2}\beta^{*} \right)\,-\left( \frac{V}{2}\right){\rm sech}\left(\frac{V}{2}\beta^{*} \right)=0$.
	In addition, $\frac{C-V}{V}-\frac{\rm sech\left(\frac{V}{2}\beta \right)}{\rm sech\left(\frac{C-V}{2}\beta \right)}$ is
	positive in $(0,\beta^{*}]$ and negative in $(\beta^{*},\infty)$,
	i.e., $T(\beta)$ is
	positive in $(0,\beta^{*}]$ and negative in $(\beta^{*},\infty)$.
	Following the same argument,
	we obtain that, 	
	if $\frac{V}{2}>\frac{C-V}{2}>0$, i.e., $\frac{C}{V}<2$, $T(\beta)$ is
	negative in $(0,\beta^{*}]$ and positive in $(\beta^{*},\infty)$.

	If $\frac{C-V}{2}>\frac{V}{2}>0$, i.e., $\frac{C}{V}>2$,
	According to Eq. \eqref{eq:refname10}, the derivative of $H(\beta)$ is positive in $(0,\beta^{*}]$ and negative in $(\beta^{*},\infty)$.
	The $H(\beta)$ is alway positive, since $\lim \limits_{\beta \to 0} H(\beta)=0$ and $\lim \limits_{\beta \to \infty} H(\beta)=0$.
	Therefore,
	when the ratio of $C$ to $V$ exceeds $2$,  $\frac{dx^{*}}{d \beta }$
	is always negative.
	Hence, $x^{*} $ is a decreasing function of $\beta$.

	Similarly, 	
	if $\frac{V}{2}>\frac{C-V}{2}>0$, i.e., $\frac{C}{V}<2$,
	the derivative of $H(\beta)$ as a function of selection intensity is
	negative in $(0,\beta^{*}]$ and positive in $(\beta^{*},\infty)$.
	Since the $\lim \limits_{\beta \to 0} H(\beta)=0$ and $\lim \limits_{\beta \to \infty} H(\beta)=0$, the $H(\beta)$ is alway negative.	
	So when the ratio of $C$ to $V$ is below 2,
	the derivative of the equilibrium, i.e., $\frac{dx^{*}}{d \beta }$, is always positive.
	Hence, $x^{*} $ is an increasing function of $\beta$.

\section{Alternative strategy update rule as selection intensity rescaling}
\label{appendixd}
We take \cite{ref41} as an example to show that strategy updating rule sometimes can be mapped to the Fermi rule with a rescaled selection intensity.
This result is fundamental to check the robustness of evolutionary rules on the final vaccination level.
Once the rescaling is made,
we make use of Section \ref{sec:selectionintensity} to reveal how different updating rules alter the final vaccination level.
	
Usually, individual updates its strategy just by comparing its payoff with the payoff of a randomly selected neighbor, which is defined as the original rule.
In \cite{ref41}, it raised a new rule where an individual compares its payoff with the average payoff who take the same strategy as its opponent.

To simplify the following analysis, we define
	\begin{equation}
	G\left[ \beta(f_{j}-f_{i})\right] =\frac{1}{1+\exp\left [ -\beta \left ( f_{j}-f_{i} \right ) \right ]}.
	\label{eq:refname19}
	\end{equation}
	
    For the original rule, the dynamics of the vaccination level can be captured by
    \begin{equation}
    \begin{aligned}
    \dot{x_{1}}=&x_{1}(1-x_{1})w(x_{1})\left(G\left[ \beta(f_{V}-f_{UI})\right]-G\left[ \beta(f_{UI}-f_{V})\right]\right) +\\
    &x_{1}(1-x_{1})(1-w(x_{1}))\left(G\left[ \beta(f_{V}-f_{UH})\right]-G\left[ \beta(f_{UH}-f_{V})\right]\right)  .
    \label{eq:refname20}
    \end{aligned}
    \end{equation}

    For the new rule, the dynamics of the vaccination level can be captured by
    \begin{equation}
    \begin{aligned}
    \dot{x_{2}}=x_{2}(1-x_{2})\left(G\left[ \beta(f_{V}-f_{U})\right]-G\left[ \beta(f_{U}-f_{V})\right]\right) ,
    \label{eq:refname21}
    \end{aligned}
    \end{equation}
    where $f_{U}=f_{UI}w(x_{2})+f_{UH}(1-w(x_{2}))=(-1)w(x_{2})+0(1-w(x_{2}))=-w(x_{2})$.

    In order to study the effect of selection intensity,
    we define $\dot{x_{1}}=\Phi_{1}(\beta)$   $\dot{x_{2}}=\Phi_{2}(\beta)$
    and perform the Taylor expansion of Eq. \eqref{eq:refname20} \eqref{eq:refname21} in the vicinity of $\beta=0$.

   \begin{equation}
     \Phi(\beta)=\Phi(0)+\Phi'(0)\beta+\frac{\Phi''(0)\beta^{2}}{2}+\frac{\Phi'''(0)\beta^{3}}{6}+o(\beta^{3}).
   \label{eq:refname22}
   \end{equation}

   For the original rule,

   $\Phi_{1}(0)=0$

   $\Phi_{1}'(0)=2x_{1}(1-x_{1})G'(0)(w(x_{1})(f_{V}-f_{UI})+
   (1-w(x_{1}))(f_{V}-f_{UH}))=2x_{1}(1-x_{1})G'(0)(w(x_{1})+f_{V})$

   $\Phi_{1}''(0)=0$

   $\Phi_{1}'''(0)=2x_{1}(1-x_{1})G'''(0)\left(w(x_{1})(f_{V}-f_{UH})^{3}+
   (1-w(x_{1}))(f_{V}-f_{UH})^{3}\right) $

   For the new rule,

   $\Phi_{2}(0)=0$

   $\Phi_{2}'(0)=2x_{2}(1-x_{2})G'(0)(f_{V}-f_{U})=2x_{2}(1-x_{2})G'(0)(f_{V}+w(x_{2}))$

   $\Phi_{2}''(0)=0$

   $\Phi_{2}'''(0)=2x_{2}(1-x_{2})G'''(0)(f_{V}-f_{U})^{3}=2x_{2}(1-x_{2})G'''(0)\left(w(x_{2})(f_{V}-f_{UI})+(1-w(x_{2}))(f_{V}-f_{UH})\right)^{3}$

   It is obvious that  $\Phi_{1}(0)=\Phi_{2}(0)$, $\Phi_{1}'(0)=\Phi_{2}'(0)$, $\Phi_{1}''(0)=\Phi_{2}''(0)$.
   Because $(x^{3})''=6x$, $g(x)=x^{3}$ is a concave function when $x>0$.
   For $\lambda \in(0,1)$, we have
   $g(\lambda x_{1}+(1-\lambda)x_{2})\leq \lambda g(x_{1})+(1-\lambda)g(x_{2}) $.
   Hence, $\left(w(x)(f_{V}-f_{UI})+(1-w(x))(f_{V}-f_{UH})\right)^{3} \leq w(x)(f_{V}-f_{UH})^{3}+(1-w(x))(f_{V}-f_{UH})^{3}$,
   i.e., $\Phi_{2}'''(0)\leq \Phi_{1}'''(0)$.

   So $\Phi_{2}(\beta)\leq \Phi_{1}(\beta)$,
   i.e., the new rule mirrors diluting the selection intensity.

\end{appendix}	

\section*{Acknowledgment}
	We acknowledge the sponsorship by the NSFC (Grants No.61603049, No.61751301).

\bibliographystyle{elsarticle-num}
\bibliography{mybibfile}
\end{document}